\documentclass[aps,prb,twocolumn,showpacs,groupedaddress]{revtex4}
\usepackage{graphicx}

\begin{document}

\title{Magnetoelastic Effects and Magnetic Phase Diagram of Multiferroic $DyMn_2O_5$}
\author{C. R. dela Cruz$^1$, B. Lorenz$^{1}$, Y. Y. Sun$^{1}$, C. W. Chu$^{1,2,3}$, S. Park$^4$, and S.-W. Cheong$^4$}
\affiliation{$^{1}$Department of Physics and TCSUH, University of
Houston, Houston, TX 77204-5002} \affiliation{$^{2}$Lawrence
Berkeley National Laboratory, 1 Cyclotron Road, Berkeley, CA 94720}
\affiliation{$^{3}$Hong Kong University of Science and Technology,
Hong Kong, China} \affiliation{$^4$Department of Physics $\&$
Astronomy and Rutgers Center for Emergent Materials, Rutgers
University, Piscataway, NJ 08854}
\date{\today }

\begin{abstract}
The magnetoelastic coupling in multiferroic $DyMn_2O_5$ is
investigated by magnetostriction measurements along the three
crystallographic orientations. Strong lattice anomalies as a
function of the magnetic field are detected at the low temperature
magnetic and ferroelectric phase transitions. The sign and magnitude
of the magnetostrictive coefficient as well as the lattice anomalies
at the transitions are correlated with the Dy moment order and with
the sharp changes of the dielectric constant and the ferroelectric
polarization. With the magnetic field applied along the c-axis a new
high-field phase has been detected the magnetic structure of which
has yet to be explored.
\end{abstract}

\maketitle











Among the multiferroic materials displaying the coexistence of
magnetic and ferroelectric (FE) orders the rare earth manganites,
$RMn_2O_5$ ($R$=rare earth, Y), have attracted attention because of
a wealth of fundamental physical phenomena observed in these
compounds. Incommensurate magnetic orders, spin frustration, lock-in
transitions, and ferroelectricity have been detected with decreasing
temperature and give rise to a cascade of phase
transitions.\cite{prellier:05} The physical origin of this phase
complexity lies in the partially competing interactions between the
$Mn^{4+}/Mn^{3+}$ spins, the rare earth magnetic moments, and the
lattice.\cite{blake:05} Geometric magnetic frustration among the
$Mn^{3+}-Mn^{4+}$ spins leads to a ground state degeneracy of the
magnetic states that can be lifted by a distortion of the lattice
(magnetic Jahn-Teller effect). This is believed to be the origin of
ferroelectricity in $RMn_2O_5$
compounds.\cite{kagomiya:05,blake:05,delacruz:06} Strong
spin-lattice coupling is needed to explain the ionic displacements
in the FE state. The experimental proof of sizable lattice anomalies
at the FE transitions in $RMn_2O_5$ was given recently by thermal
expansion experiments in zero magnetic field.\cite{delacruz:06}
However, the effect of an external magnetic field on the lattice of
$RMn_2O_5$ has not been investigated yet. The field couples to the
magnetic order resulting in field-induced spin reorientations and
magnetic phase transitions\cite{ratcliff:05} which, in turn, should
affect the lattice via the spin-lattice interaction. The signature
and the nature of the magnetoeleastic effect provides essential
information about the intrinsic magnetoelastic and magnetoelectric
interactions. We have therefore investigated the magnetoelastic
strain on the lattice of $DyMn_2O_5$ and found large changes of the
lattice parameters as a function of the magnetic field and abrupt
striction effects at phase boundaries.

$DyMn_2O_5$ exhibits a complex magnetic phase
diagram.\cite{ratcliff:05, higashiyama:04} Large effects of an
$a$-axis magnetic field on the dielectric constant\cite{hur:04} and
the FE polarization\cite{higashiyama:04} have been reported. Upon
decreasing temperature $T$ the sequence of phase transitions is
similar for most $RMn_2O_5$. AFM order develops at $T_{N1}=43$ K
with an incommensurate (IC) magnetic modulation described by
$\overrightarrow{q}=(0.5+\beta,0,0.25+\delta)$ (where
$\beta,\delta<0$ indicate the deviations from commensurability)
followed by a lock-in transition at $T_{C1}=39$ K into a
commensurate FE phase with $\overrightarrow{q}=(0.5,0,0.25)$. At the
low-$T$ side a transition into a re-entrant IC magnetic phase with
modulation vector $\overrightarrow{q}\approx(0.48,0,0.3)$ was
reported for all $RMn_2O_5$ ($T_{C3}$=6.5 K in $DyMn_2O_5$).
$DyMn_2O_5$ experiences two more phase changes in between $T_{C1}$
and $T_{C3}$, a spin reorientation at $T_{N2}=27$ K and a lock-in
transition accompanied by another spin reorientation at $T_{C2}=13$
K.\cite{ratcliff:05} Fig. 1 shows the dielectric constant upon
heating with the phase transition temperatures and the different
phases labeled as used throughout this communication. All five phase
transitions are clearly visible as distinct anomalies of the
dielectric constant, the FE polarization,\cite{higashiyama:04} the
heat capacity, and the lattice parameters.\cite{delacruz:06} The
strongest lattice anomalies have been observed at $T_{C3}$ where the
Dy moments undergo the AFM odering with
$\overrightarrow{q}_{Dy}=(0.5,0,0)$ and the FE polarization drops to
a small value close to zero. This reentrant paraelectric phase and
its stability with respect to magnetic fields is of particular
interest.

Single crystals of $DyMn_2O_5$ have been grown as described
elsewhere.\cite{hur:04} The longitudinal magnetostriction
measurements were conducted along the three crystallographic
orientations in an Oxford variable temperature cryostat with a
superconducting magnet up to 14 T employing a high-precision
capacitance dilatometer.\cite{delacruz:06} Complimentary
measurements of the $b$-axis dielectric constant in magnetic field
have been conducted to correlate the observed field-induced lattice
anomalies with the dielectric properties and the known phase
transitions. The dielectric constant was derived from the
capacitance of a thin platelet with silver electrodes measured by
the HP4285A LCZ meter at 1 MHz.

Fig. 2a shows $\Delta a/a$ increasing with the magnetic field,
$H_a$, at 6 K. The sharp step at $H_{C3}=1.44$ T indicates the first
order phase transition from the "PE" phase to the "FE3" phase with a
field hysteresis of 0.85 T. The inset of Fig. 2a shows the
derivative (magnetostrictive coefficient) and the sharp peak-like
anomalies with increasing and decreasing field at $H_{C3}$. The
field hysteresis is in accordance with the temperature hysteresis at
$T_{C3}$ observed in various physical quantities. The phase boundary
between the "PE" and "FE3" phases (Fig. 2d) is derived from the
lattice anomalies at different temperatures. The dashed area
designates the hysteretic region in the phase diagram. The phase
boundary is in good agreement with the phase diagram derived from
polarization measurements\cite{higashiyama:04} and from neutron
scattering experiments.\cite{ratcliff:05} No additional lattice
anomalies could be observed up to 13 T. The previously observed
field-induced transitions from "FE3" to "FE2"\cite{higashiyama:04}
or into more complex structures\cite{ratcliff:05} are apparently not
accompanied by sizable changes of the $a$-axis length. However,
above $H_{C3}$ we can identify a linear magnetoelastic region
extending to about 3 T (dashed line in Fig. 2a), similar to the
field-induced linear magnetoelastic effect discovered in
$TbMnO_3$.\cite{aliouane:06} In this field range the growth of a
ferromagnetic (FM) moment and a change of the polarization extending
to about 4 T (the maximum of $a(H)$) has been
observed.\cite{ratcliff:05} It appears conceivable that the linear
magnetoelastic region and the following maximum of $a(H)$ are
related to the commensurate FM and the incommensurate FM phases with
a supposedly canted spin structure suggested by neutron scattering
experiments.

The expansion of the $a$-axis with the field has its origin in the
release of the strong AFM interaction of the Dy. The magnetic
structure of the "PE" phase is dominated by the AFM order of the Dy
moments along $a$, $\overrightarrow{q}_{Dy}=(0.5,0,0)$. The magnetic
system can gain exchange energy by reducing the distance of the
ordered Dy moments along $a$ resulting in the large and sudden
contraction of the $a$-axis below $T_{C3}$.\cite{delacruz:06} The
external field $H_a$ reduces the AFM Dy order along this axis and
partially releases the lattice strain causing the increase of $a$ at
small $H_a$. At the critical field $H_{C3}$ neutron scattering
experiments have indicated a reorientation of the Dy moments and a
sudden reduction of their mutual correlations. This results in the
sharp increase of $a$ at $H_{C3}$ (Fig. 2a). Our results provide
clear evidence for the strong coupling of the Dy moments to the
lattice.

The length of the $b$-axis decreases with increasing $H_b$ (Fig.
2b). The transition from "PE" to "FE3" is clearly seen as a sharp
drop of $b$ at $H_{C3}$. At higher fields a second step-like
decrease of $b$ at $H_{C2}$ indicates the field-induced transition
into the "FE2" phase. Both phase transitions exhibit a distinct
field hysteresis in accordance with the temperature hysteresis at
$T_{C3}$ and $T_{C2}$ observed at zero field. The data shown in Fig.
2b at 7 K fall into the temperature hysteretic region close to
$T_{C3}$ and the reverse transition from "FE3" to "PE" does not take
place in decreasing field. The phase sequence at 7 K with cycling
the magnetic field from 0 to 11 T and back is therefore
"PE"($H_b$=0) $\Rightarrow$ "FE3"($H_b>$1 T) $\Rightarrow$
"FE2"($H_b$=11 T) $\Rightarrow$ "FE3"($H_b$=0). Above 8 K the
low-field anomaly of $b$ disappears and the high-field drop of
$b(H)$ can be traced to $T>$15 K. The phase diagram is shown in Fig.
2e. The shaded areas designate the hysteretic regions. Both phase
boundaries match the zero-field transition temperatures $T_{C3}$ and
$T_{C2}$. The stability of the "FE3" phase extends to far higher
fields as in the case of $H\|a$.\cite{higashiyama:04, ratcliff:05}
The opposite sign of the magnetostriction along $b$ as compared to
$a$ is consistent with the anisotropy of the thermal expansion
anomalies observed at $T_{C3}$.\cite{delacruz:06} The
low-temperature "PE" phase shows a sizable reduction of the $a$-axis
length but a large increase of $b$ and $c$. The shrinkage of $a$ is
driven by the $Dy$-moment order along $a$ as discussed above. The
expansion of $b$ and $c$ below $T_{C3}$ is then a consequence of the
elastic forces of the lattice. The external magnetic field applied
along all directions destabilizes the "PE" phase and releases the
magnetostrictive strain. This results in the increase of $a$ but a
decrease of $b$ and $c$ with the applied field.

The $c$-axis of $DyMn_2O_5$ is the hard magnetic axis\cite{hur:04}
and the magnetic phase diagram for $H_c$ has not been investigated
yet. The magnetostriction measurements along $c$ reveal a more
complex phase diagram as compared to the $a$- and $b$-axis fields.
At low $T$ a sharp drop of $c(H)$ at about 4 T indicates a phase
transition with a field hysteresis of up to 1 T. Magnetostriction
data at different temperatures are shown in Fig. 2c. The phase
boundary of the "PE" phase shifts to lower fields with increasing
temperature and vanishes above 7.8 K. At higher $T$, however,
another magnetostrictive anomaly above 4 T indicates a phase
transition into a high-field phase ("HF"). The critical field of
this transition increases with increasing $T$ (Fig. 2f). The
transition "FE3" $\Leftrightarrow$ "HF" exhibits a much larger field
hysteresis of up to 3.5 T (as can be seen from the 8.5 K data of
Fig. 2c and the phase diagram of Fig. 2f). The magnetic structure of
the "HF" phase is yet to be explored. It appears conceivable that
the magnetic field realigns the Dy moments resulting in a spin-flop
transition similar to that observed in $GdMnO_3$.\cite{hemberger:04}
The $c$-axis magnetization (inset to Fig. 2c) reveals a metamagnetic
transition at this phase boundary that is consistent with a
spin-flop scenario. The critical field indicating the entrance into
the "HF" phase depends on the starting phase at zero field. For
temperatures below $T_{C3}$ the transition "PE" $\Rightarrow$ "HF"
happens below 4 T and is associated with a larger step of $c(H)$
(data at 4.8 and 7 K in Fig. 2c) whereas the transition from "FE3"
to "HF" proceeds above 4 T with a significantly smaller change of
the $c$-axis upon increasing field. Due to the wide hysteresis of
the transitions there is a large region in the phase diagram
(indicated by horizontal dashed lines) in which a strong
($T,H$)-history dependence of the final state can be observed. An
example is given in Fig. 3a. At a temperature of 7 K both phases,
"PE" and "FE3", can be stabilized depending on approaching this
temperature from below or above, respectively. Cooling to 7 K
preserves the system in the "FE3" phase. Increasing the magnetic
field from zero to above 5 T transforms the system into the "HF"
phase. The subsequent release of the field causes a phase transition
from the "HF" phase into the "PE" phase at about 1 T. The $c$-axis
lattice parameter shows the sharp drop at the "FE3" $\Rightarrow$
"HF" transition with increasing $H$ and a large increase at the "HF"
$\Rightarrow$ "PE" transition with the release of $H$ to zero. After
the field cycle, in the final state, $c$ is larger than in the
initial state. The difference of $\Delta c/c\approx6\cdot10^{-5}$ is
in perfect agreement with the lattice strain across $T_{C3}$
obtained from zero-field thermal expansion
experiments.\cite{delacruz:06} Due to the thermal hysteresis region
around 7 K at $H=0$ the external field can "switch" the magnetic and
FE states and the $c$-axis length as indicated in Fig. 3a.

The phase diagram of Fig. 2f has been verified by measurements of
the dielectric constant, $\varepsilon (T,H)$, in magnetic fields up
to 7 T. The various phase transitions between the "PE", "FE3", and
"HF" phases are clearly detected in sharp changes of $\varepsilon$
upon increasing and decreasing fields. Some examples of $\varepsilon
(H)$ are shown in Figs. 3b to 3d. The phase boundaries derived from
dielectric measurements agree perfectly with the ones from
magnetostriction data and they are included as stars in Fig. 2f.
Moreover, the irreversible phase change "FE3" $\Rightarrow$ "HF"
$\Rightarrow$ "PE" while cycling $H$ at 7 K is also verified from
the dielectric data (included as stars in Fig. 3a). There is a small
region in the phase diagram between 7 K and 7.8 K where the increase
of $H_c$, starting from the low-temperature "PE" phase, results in
two subsequent phase transitions, first into the "FE3" phase and
then into the "HF" phase. Fig. 3c shows $\varepsilon (H)$ at 7.5 K.
With increasing field the sharp increase of $\varepsilon$ is due to
the transition "PE" $\Rightarrow$ "FE3" whereas the sudden drop at
higher field signals the second transition "FE3" $\Rightarrow$ "HF".
Upon decreasing field the large hysteresis prevents any further
phase change (Fig. 3c).

The major phase boundaries of Fig. 2f have been derived from
magnetostriction and dielectric constant measurements at constant
temperature with changing magnetic field. In order to uniquely
identify the stability regions of the different phases temperature
dependent measurements of $\varepsilon$ at constant field have been
conducted. The major phase boundaries were found in agreement with
the field dependent measurements, however, the data have revealed an
additional hysteretic region below 7 K close to 4 T. This area is
marked by vertical dashes in Fig. 2f. With the field and temperature
dependent measurements of $c(T,H)$ and $\varepsilon(T,H)$ we can
identify the thermodynamically stable single phase regions of the
phase diagram (shown by the blank areas in Fig. 2f). In between
these stable phase regions two or more phases can coexist and the
magnetic and FE states of $DyMn_2O_5$ are history dependent.

The new "HF" phase detected in magnetic fields along the $c$-axis
and the strong lattice anomalies across the phase boundaries
indicate that external fields have a significant effect not only if
they are aligned with the axis of easy magnetization $a$ but also
along the hard magnetic axis $c$, contributing to an increased phase
complexity in $DyMn_2O_5$. This is not surprising since the magnetic
modulation of the Mn and Dy order has an important component along
the $c$-axis resulting in a spiral order of moments along $c$ which,
depending on the temperature range, can be commensurate ($q_z=0.25$)
or incommensurate ($q_z=0.25+\delta$).\cite{ratcliff:05} Recent
investigations of the magnetic phase diagram of $DyMn_2O_5$ have
focused on $H_a$ and the magnetic structures have largely been
resolved for this case.\cite{higashiyama:04,ratcliff:05} However,
the field-induced "HF" phase ($H_c$) and the magnetic order of
$Dy$-moments and $Mn$-spins in this phase have yet to be explored.
Neutron scattering experiments in magnetic fields $H_c$ are
therefore highly desired to resolve the complex magnetic structures
in the various phases.

It has long been speculated that magnetoelastic interactions play a
major role and are the key to understand the magnetically induced
ferroelectricity and the complex phase diagrams of multiferroic
$RMn_2O_5$ compounds. Our results provide the experimental prove of
the existence of strong magnetoelastic couplings in $DyMn_2O_5$. The
sharp lattice anomalies observed at the phase transitions induced by
temperature\cite{delacruz:06} or by magnetic fields (Fig. 2) reveal
the intrinsic spin-lattice coupling. Any change of the magnetic
order results in a visible strain on the lattice. In addition, there
is a sizable coupling of the lattice to an external field within a
single phase. For example, the total change of the $c$-axis length
in fields up to 8 T is $3\cdot10^{-4}$, far larger than the
magnetostriction along $a$ or $b$. This indicates a high sensitivity
of the lattice with respect to $H_c$ although the magnetization
change in this direction is the smallest one.\cite{hur:04} The
magnetic orders and the ferroelectricity in multiferroic compounds
are highly correlated. The effects of magnetostriction and
multiferroic coupling cannot be separated. The current results show
the macroscopic strain the lattice experiences as a result of
magnetic / ferroelectric phase changes and the action of an external
magnetic field. Lattice anomalies across phase transitions that are
associated with a major change of the ferroelectric polarization
(e.g. the transition from "PE" to "FE3" phase) could be ascribed to
multiferroic coupling effects but the magnetic order also
experiences a change at this transition. The anisotropy of the
magnetostriction along the different axes provides additional
insight into the magnetoelastic interactions and, in the case of
$DyMn_2O_5$, strongly suggests the dominant role of the $Dy$-moment
order at lower temperatures.

\begin{acknowledgments}
This work is supported in part by the T.L.L. Temple Foundation, the
J. J. and R. Moores Endowment, and the State of Texas through the
TCSUH and at LBNL by the DoE. The work at Rutgers is supported by
NSF-DMR-0520471.
\end{acknowledgments}

\bibliographystyle{phpf}           

\newpage

\begin{figure}
\caption{(Color Online) Dielectric constant of $DyMn_2O_5$ with all
anomalies indicating transitions between different magnetic and FE
structures. Data shown are for increasing $T$.}

\caption{(Color Online) (a) to (c): Longitudinal magnetostriction
along the $a$-, $b$-, and $c$-axis of $DyMn_2O_5$ (Data in the main
panels of (a) and (b) are shown for increasing field only, the
insets show the derivative for both field directions.) The inset to
(c), c-axis magnetization vs. field, show the metamagnetic
transition at the phase boundary "PE" $\rightarrow$ "HF". \newline
(d) to (f): Derived low-temperature phase diagrams. The regions of
field hysteresis are indicated by dashed lines.}

\caption{(Color Online) (a): Magnetostriction of the $c$-axis
(circles, left scale) and dielectric constant (stars, right scale)
at 7 K showing the irreversible change from "FE3" to "PE" induced by
the magnetic field cycle. (b) to (d): Dielectric constant at three
characteristic temperatures showing sharp anomalies at all magnetic
transitions between the "PE", "FE3", and "HF" phases. The directions
of field change are indicated by arrows.}
\end{figure}

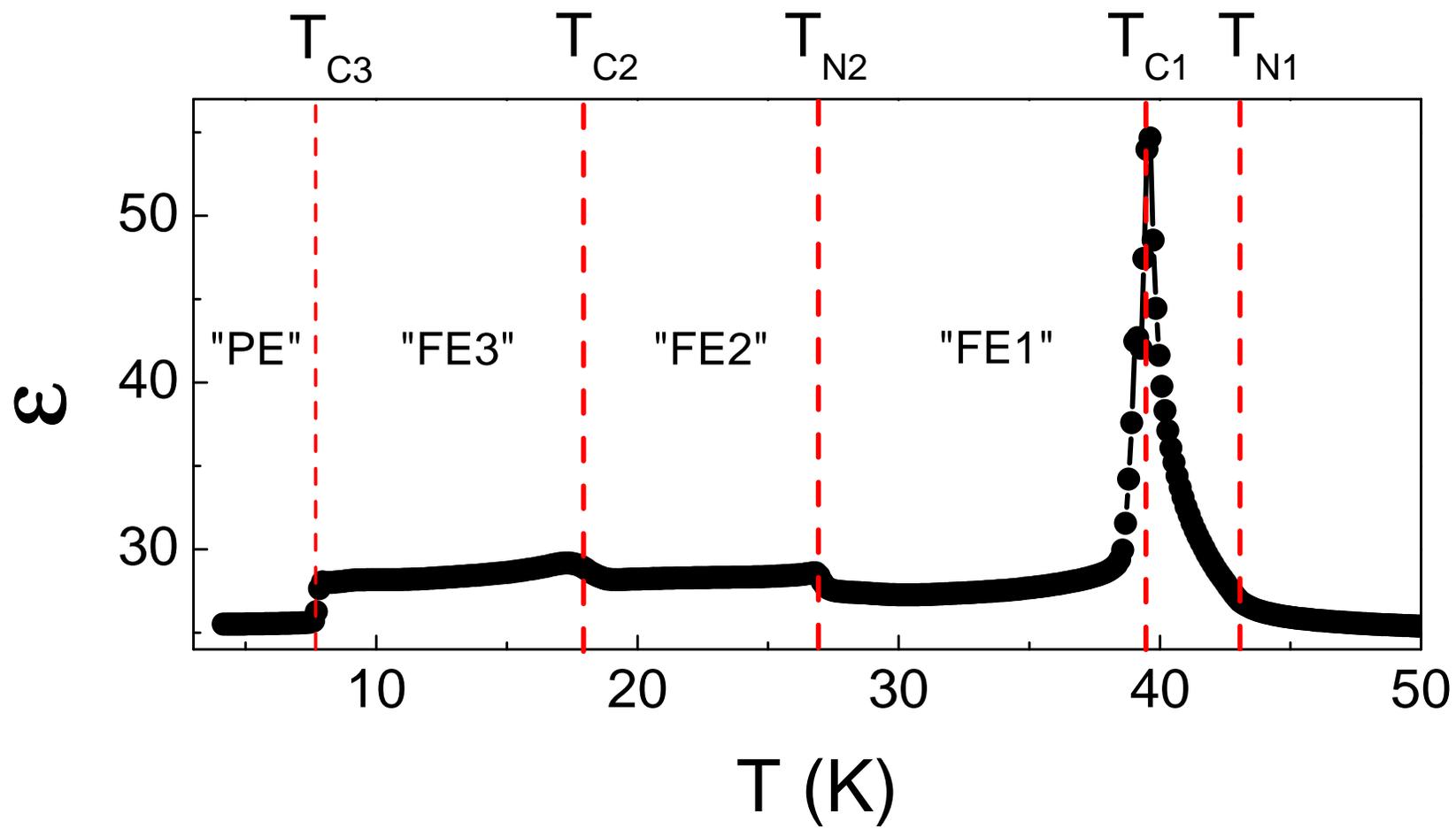
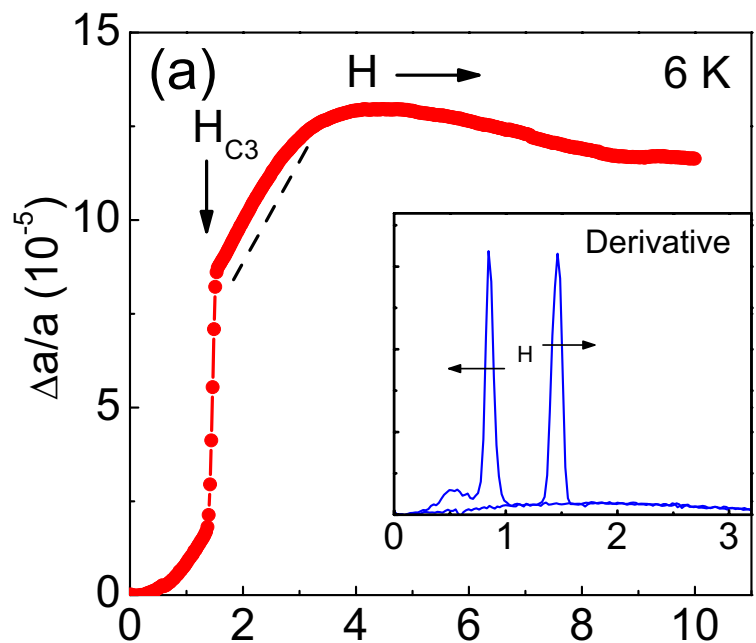
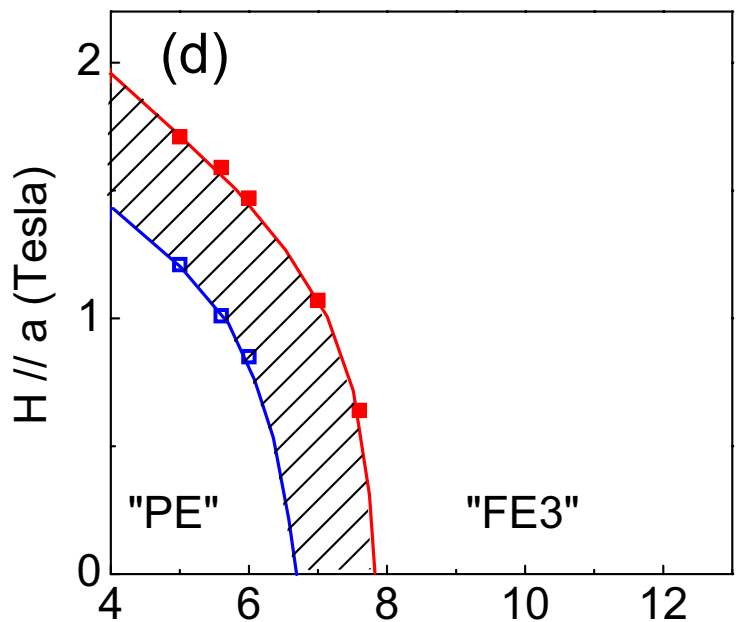
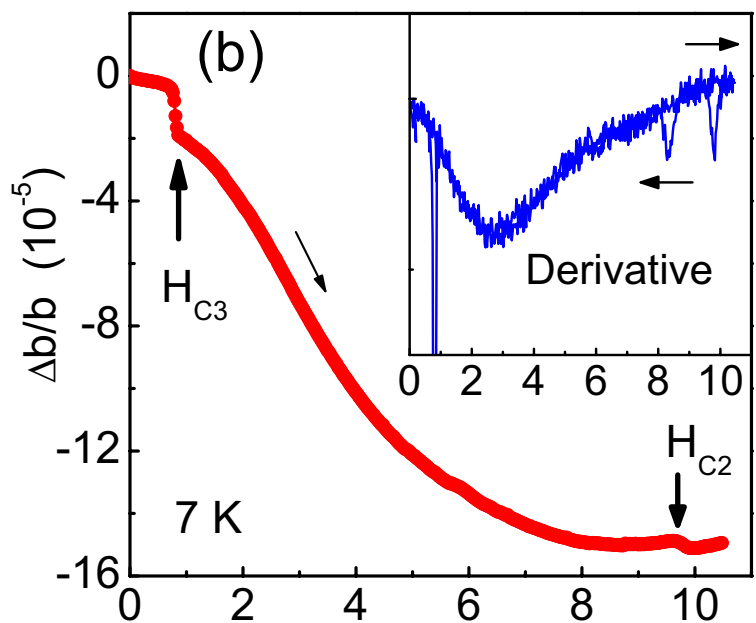
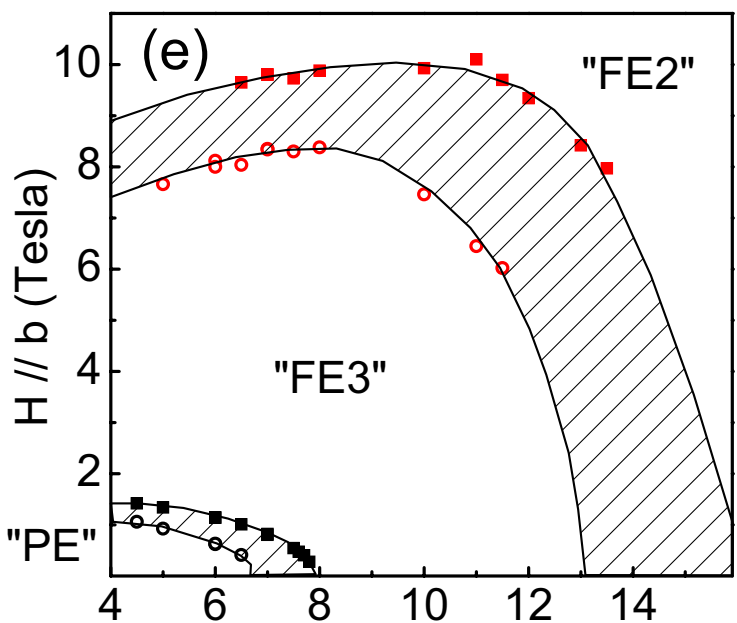
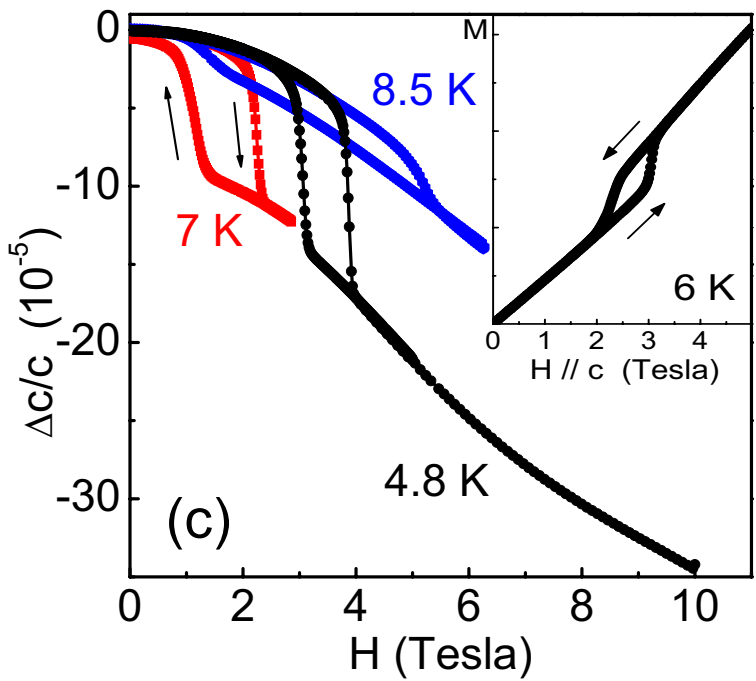
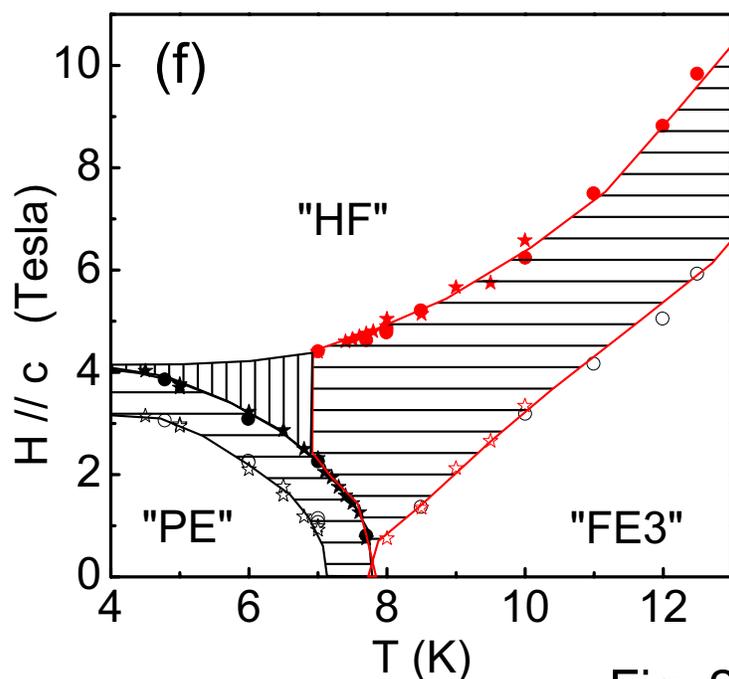
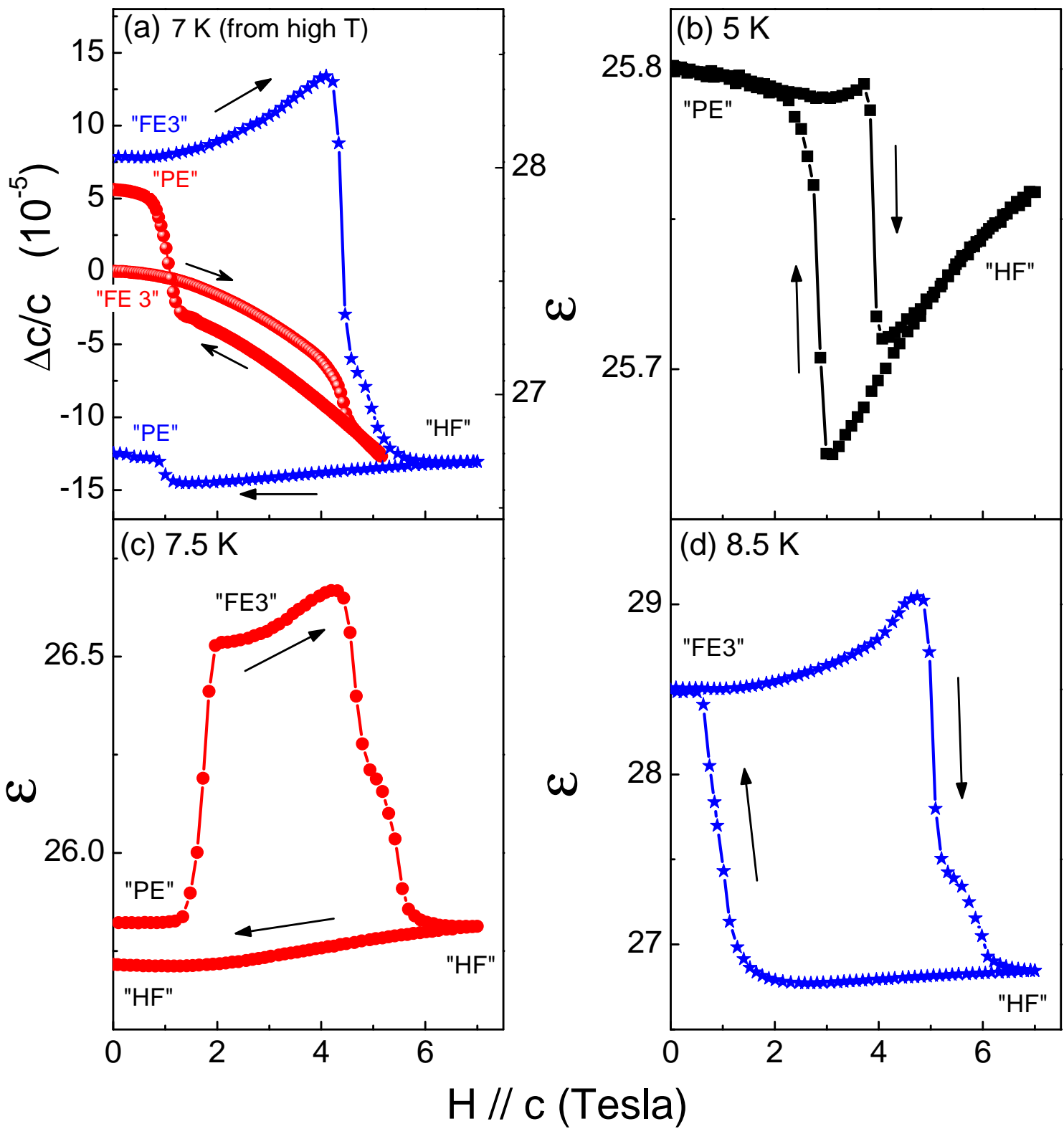
\end{document}